\documentstyle[11pt]{article}
\begin{document}
\title{Universe as a Phase Boundary in a Four-Dimensional Euclidean
Space}
\author{Michael Grady\\
Department of Physics, SUNY College at Fredonia,
Fredonia NY 14063 USA}
\date{\today}
\maketitle
\thispagestyle{empty}
\begin{abstract}
It is proposed that space is a four-dimensional Euclidean space with
universal time. Originally this space was filled with a uniform
substance, pictured as a liquid,
which at some time became supercooled.
Our universe began as a nucleation event initiating
a liquid to solid transition. The universe we
inhabit and are directly aware of consists of only the
three-dimensional
expanding phase boundary. Random energy
transfers to the boundary from thermal fluctuations
in the adjacent bulk phases are interpreted by us as quantum
fluctuations.
Fermionic matter is modeled as screw dislocations;
gauge bosons as phonons.  Minkowski space emerges
dynamically through redefining local time to be proportional
to the spatial coordinate
perpendicular to the boundary. Other features include
a geometrical quantum gravitational theory, and an
explanation of quantum measurement.
\end{abstract}
\newpage
In the following, a new picture of the big bang and the underlying
structure of the universe is proposed, based on a classical field
theory in four-dimensional Euclidean space with a universal time
(a 4+1 dimensional theory). The big bang is treated as a
nucleation event
for a first-order phase transition (pictured as a liquid to
solid transition)
and our universe is the three-dimensional phase boundary
between the expanding
solid and preexisting liquid phases. This classical theory
appears to be
able to explain a diverse set of phenomena -- the expansion
of the universe
at a non-decreasing rate, special relativity (which arises dynamically
from the expanding phase boundary), quantum fluctuations in terms of
four-dimensional thermal fluctuations, quantum measurement in terms of
classical spontaneous symmetry breaking, and a quantum
theory of gravity
based on the geometry of the expanding hypersurface.
The specific model of a growing crystal
allows one to model elementary fermions
as screw
dislocations and bosons as phonons.
Collisions with other universes provides  possible
explanations for the
pattern of matter distribution in the universe and for
the existence of quasars.

We begin by assuming a four-dimensional
Euclidean space, filled with a uniform fluid at some temperature,
undergoing thermal fluctuations. In addition to the four spatial
dimensions, there is also a universal time.  Another possibility would
be to start with a five-dimensional Minkowski space.
This liquid was cooling,
became supercooled, and at some point a solid crystal nucleated.
This was the big bang. The universe begins as a
fluctuation, already at a finite size, because in order to grow
rather than shrink, the initial crystal must be large enough
that the positive surface energy is less than the negative
volume energy
relative to the liquid. The {\em surface} of the solid,
the phase boundary, is
an expanding three-dimensional space, our universe.
This differs from other ``bubble universe'' pictures, where the
universe is the {\em interior} of a {\em 3-d} bubble.
We are not directly aware of the relatively uniform liquid and solid
phases, but only of the phase boundary between them,
which we refer to as the ``present''.  As the
crystal grows, this hypersurface,
our universe, expands. Already there
is a variance with the usual $\Lambda=0$ Friedmann universes.
Namely, our universe is closed, but will expand forever. The pressure
on the surface caused by the energy difference of the two
phases acts like a repulsive cosmological constant.
This universe actually expands faster as time goes on, not slower. If
dissipation is present it will eventually
approach a constant rate.
(This assumes a constant amount of supercooling -- if the
base liquid cools
more, the
expansion rate could continue to
increase as the degree of supercooling
increases. Without dissipation, the expansion rate
increases exponentially).
Recent astrophysical evidence shows that the expansion rate is not
slowing,
but may even be speeding up\cite{expand} which is consistent
with this scenario.

   The basic theory needed to describe this
expanding phase boundary is non-equilibrium classical
statistical mechanics.
The solid, in some sense, lies in the past, since
we have been there earlier, although it still exists
in the present when
observed from the higher dimension.  The liquid represents
the future, since that is where we are going, but it also exists
now, as an undifferentiated, fluctuating medium. To distinguish the
current states of the solid and the liquid from our own
past and future,
they may be called the ``current past'' and ``current future''.
They differ from our past and future because changes
may have occurred
after the solid was formed, and the future certainly will be
different when we
arrive there.
To the extent that the solid is frozen, however,
our past may be accurately preserved within it.
We may not be aware of the existence of the
liquid due to its uniformity.
However, the boundary which we inhabit is in thermal contact
with both the liquid and solid phases, and can
certainly exchange energy
with them. Thus objects riding the interface
will get random energy
fluctuations from this thermal contact. These
random thermal fluctuations
could explain quantum fluctuations. It is well known
that in ordinary quantum theory, if Minkowski
space is analytically continued
to Euclidean space, quantum fluctuations behave
as higher-dimensional thermal fluctuations, i.e. the Feynman
path integral becomes an ordinary statistical mechanical
partition function
in 4 (+1) dimensions.
Plank's constant is proportional to the temperature of
the four-dimensional
Euclidean space.
In such a picture a quantum phenomenon such as tunneling is explained
classically as due to a random kick of extra energy which results
from thermal contact with the liquid and solid phases.  Due to such
thermal fluctuations, energy is not conserved over short time periods;
it is conserved only in the average over time.

Usually, analytic continuation to
Euclidean space is seen as a mathematical trick to transform the
poorly-defined physical Minkowski-space theory into a
Euclidean-space partition function that is easier to handle.
Here, it is proposed that Euclidean space {\em is} the
correct physical
space. It is Minkowski space which results from
a mathematical trick designed to describe a decidedly non-equilibrium
feature, the expanding phase boundary,
within the formalism of equilibrium
statistical mechanics.
In single-phase equilibrium statistical mechanics, correlation
functions decay exponentially with distance or time.
However, a feature like an expanding phase boundary propagates without
decaying. One way to represent such a feature is to give it an
imaginary energy, which produces a pole in the Fourier
transform of the
correlation function.
To keep time as the Fourier conjugate of energy,
it must also be taken imaginary. In this way, nondecaying features
can be modeled  within equilibrium
statistical mechanics, by just the inverse
of the trick often used to allow particle decay in quantum mechanics.
This ``mathematical trick'' results in a Minkowski space
in the ``$ict$'' formulation.
If observers in the hypersurface choose their
local time coordinate to be proportional to the
fourth spatial coordinate
perpendicular to their expanding 3-d hypersurface, special relativity
can be fully realized. This time is equal to
the  product of universal
time and the expansion rate
for observers co-moving with the expansion.

Since the background theory is a classical field theory undergoing a
phase transition one does not have to use
methods borrowed from equilibrium physics -- in fact it
is not entirely correct to do so.  A purely non-equilibrium
approach
involving, say, the Langevin equation, may be better.
The universe,
evolving with universal time, is in a definite
state at any time. However, 4-d
thermal fluctuations may create a kind
of zitterbewegung -- very rapid variation at small scales, that
enforces the uncertainty principle.
This more classical evolution
affords the opportunity to
explain the quantum measurement process
as a spontaneous symmetry breaking event\cite{ssb}.
A measuring device,
originally with an unbroken symmetry, couples
to a system under study becoming strongly
correlated with it. Then
an adjustment is made to the potential of the
measuring device which initiates spontaneous symmetry
breaking.
The measurement takes place at this time,
when the ensemble
of possible future states of the combined system splits
into non-ergodic subensembles corresponding to the
possible values of the order parameter, also corresponding
to possible values of the measured quantity.
Future evolution is confined to a single subensemble
in the usual manner of a classical symmetry-breaking phase transition.
In this picture measurements are well defined, the
collapse is a physical event, and a clear distinction exists
between what constitutes a measuring device and what does not.
Because the ``current past'' is continuously undergoing 4-d thermal
fluctuations, it is only frozen to the extent that the ensemble is
limited due to spontaneous symmetry breaking. Thus questions such as
``which slit did the electron go through'' or ``which direction was
the spin pointing'' are as meaningless here as they are in standard
quantum mechanics. This is because the details of history
are continuously
being rewritten as the current past fluctuates. Only to
the extent that the ensemble is limited by spontaneous
symmetry breaking
can one make definite statements about past events.

Using a crystal as the analogy for the
``current past'' phase affords the possibility of describing
fermions as screw dislocations.
These obey
an exclusion principle and have long range
forces with left/right-hand acting like particle/antiparticle.
They can annihilate
or be pair produced.
Screw dislocations in an ordinary three-dimensional crystal
form line defects.
In a four-dimensional crystal such dislocations have a sheet
topology, thus the basic material entity would
actually be a fermionic string.
The embedding of such an object into the crystal is rather
complex, and
may contain enough structure to explain both spin and isospin.
This is
because the O(4) rotational symmetry of 4-d space can be written as
SU(2) $\times$ SU(2).
The interface could also
allow (or actually require) the use of
domain-wall fermions, one method of obtaining
a chiral theory\cite{kaplan}.
It should also be pointed out that dislocations are a wave phenomenon;
essentially they are soliton-like displacement waves in the underlying
crystal. Thus this can also be seen as a realization of the general
idea of explaining fermions as solitons. Because they are waves, matter
can and does exhibit interference phenomena in this theory.

Of course the solid-state analogy for the photon is the phonon.
Phonons obey a relativistic-like dispersion relation,
relative to the speed of sound. One can have
phonons which travel only on the surface,
as well as within the bulk phases.
If sound speed in the crystal is to be equated with
light speed, however,
what happens to the faster-than-light prohibition of
special relativity? It actually still holds for dislocations!
It has been shown that screw dislocations cannot move faster
than the speed of sound in the medium.
Their
effective mass from the stress in the surrounding crystal grows
with speed, becoming
infinite as the speed of sound is approached, in an exact mathematical
analogy with the relativistic mass increase\cite{dislocations}.
These dislocations also experience a Lorentz-like length contraction
relative to the speed of sound, also following exactly the mathematics
of a Lorentz transformation.
This curious mathematical analogy between the behavior of dislocations
in crystals and the special theory of relativity was first noted by
Frenkel and Kontorowa in 1938\cite{dislocations}.
What is being suggested here is that
this is perhaps not just an analogy, but the actual
explanation for the
Lorentz transformation. This is reminiscent of the original
viewpoint of
Lorentz and Fitzgerald, that the contraction is a dynamical
physical effect.
If a universal time is used, then one
would have an ether theory with a preferred frame, the rest frame
of the medium, in contradiction with the Michelson-Morley experiment.
The Lorentz transformation would
work in only one direction, because it would be missing the time part
of the transformation.
However if each observer chooses a time coordinate along
their own world
line, a fully reciprocal special relativity based
on the speed of sound
as limiting velocity can arise.
This is because the operation of any moving
clock constructed from matter
will be affected by the length-contraction and mass increase
effects in such a way that it runs slow compared to a stationary clock.
In other words, the usual relativistic time dilation can be derived as
a consequence of the length-contraction and mass increase (for
a specific
model clock one can take a mass-spring oscillator, a ``light clock''
consisting of light bouncing between mirrors constructed with material
spacers, or other simple
physical systems).
The combination
of a time-dilated clock and a tilted time axis (along the observer's
world line) results in the correct Lorentz time transformation. Thus
if the moving observer chooses this as the time coordinate, a theory
exactly equivalent to special relativity arises, and the effects of
moving with respect to the ether are effectively hidden from the
moving observer (of course in our material world there are methods
even in standard special relativity
of determining the preferred frame at rest with the expansion - it
is the frame in which the
temperature of the cosmic background radiation is isotropic).

In this theory, special relativity is a dynamical effect, not a
property of the underlying space, but of the {\em particular}
solution of
an expanding phase boundary, together with our identification of
the passage of time with our motion in the fourth spatial direction
as a result of the expansion.
The usual logic of special relativity is turned around.
Length-contraction
and mass-increase are the basic phenomena, from which time-dilation,
the full Lorentz transformation, and finally the apparent constancy
of the speed of light for all inertial observers
(the usual main postulate of special relativity)
is derived (this last as a
consequence of the full Lorentz transformation).

A more geometrical, but less complete argument for the origin of
Minkowski space is as follows.
Two
observers
share a common Euclidean metric in the background space,
\[ds^{2} = dx_{1}^{2} +dx_{2}^{2} +dx_{3}^{2} +dx_{4}^{2}
= ds^{\prime\, 2} =
dx^{\prime\, 2}_{1}+dx^{\prime\, 2}_{2} +dx^{\prime\, 2}_{3}
+dx^{\prime\, 2}_{4}.\]
The direction $x_4$ is the
expansion direction for the unprimed observer, $x'_4$ that of the
primed observer (i.e. along that observer's direction of motion).
Assume each analytically continues their fourth
coordinate to imaginary
values, $t=i x_4$ and $t'=ix'_4$,
in order to model the non-decaying phase front as explained earlier.
Then the metric
becomes
\[ds^{2} = dx_{1}^{2} +dx_{2}^{2} +dx_{3}^{2} -dt^{2}
= ds^{\prime\, 2} =
dx^{\prime\, 2}_{1}+dx^{\prime\, 2}_{2} +dx^{\prime\, 2}_{3}
-dt^{\prime\, 2}.\]
and the observers
feel they are living in a 3+1 dimensional Minkowski space.

Although material objects cannot exceed the speed of sound,
the phase front
itself can expand faster than the speed of sound, as in a detonation.
This may be
necessary to isolate us from waves sent into the past that reflect
back, which does not seem to accord with experience.

Variations in the geometry of the expanding hypersurface lead to the
possibility of a general-relativity like gravitational theory,
but which
would naturally include quantum fluctuations as
4-d thermal fluctuations.
Interestingly, crystal growth is faster near dislocations, which
would distort the local surface geometry in regions where
dislocations are
concentrated; matter could affect curvature by this mechanism.
This theory would differ from general relativity, but
might be approximated by it.
It is interesting to speculate what black holes
would look like in this
picture.  They would be stalagmite-like protuberances
from the base crystal,
with sides sloped at an angle greater than 45$^{\circ}$. Surface
phonons would be non-propagating due to an expansion
rate faster than the
speed of sound. But other than this, the interior of the black hole
would not be so different from the exterior, and there would not seem
to be a need for a singularity.

It is also interesting to consider the possibility  of
collisions between two such universes.
This can be envisioned as similar to two soap bubbles colliding and
then coalescing, except that it needs to be pictured in one
more dimension.
The intersection of two
three-dimensional surfaces
is a two-dimensional surface. The ``grain boundary'' that would form
at the intersection would be full of dislocations.
This could explain
why matter in the universe appears to exist mostly in 2-d wall-like
features. In this scenario, most matter would not have been created in
the initial big bang, but in one or more cosmic
collisions shortly thereafter with other growing universes
that nucleated
nearby.
Matter would never have existed in a homogeneous distribution,
but rather
would be created in a clumpy distribution.
This would make galaxy formation
easier, but might not be able to adequately explain the
uniformity of the
cosmic background radiation.
If the nucleation events were heterogeneous, it is possible to
have a number of correlated nucleations to occur near one
another in space
and time, but be rare in other regions. After an initial series
of collisions,
the coalesced universes can then grow undisturbed into a
quieter region
of the pure liquid phase. This could explain why such
collisions do not
appear to be occurring today. Collision of our universe with a smaller
universe would be quite spectacular. One would see a rapidly expanding
shell, the collision boundary, which would at first expand
at superluminal
velocity (this is due to the geometry of the collision - even
the join-radius
of two fast-moving colliding soap bubbles could expand
superluminally because
no actual matter is travelling
superluminally).
This would also be a locus of matter formation and emit
copious radiation.
If the colliding universe were also rotating with respect
to ours, then
one handedness of dislocations would predominate over the
other, giving
a possible reason for the preponderance of matter over antimatter.
That part of our universe that existed previously
within the shell would be utterly destroyed (as would
the corresponding
part of the other universe). These regions would no longer reside on
the surface, but would now be in the interior of the crystal, buried
in the past. What would finally exist inside the expanding
spherical shell
(which would eventually slow down and stop) would be the entire other
universe, patched into our own. It is interesting to speculate that
quasars could be the result of such collisions with other
small universes. Many quasars appear to
have superluminal velocities or correlations within them\cite{quasar},
though conventional explanations may be able to explain these as
essentially optical illusions.

The expanding phase-boundary
model also has good explanations for the horizon and flatness
problems. Since the universe presumably preexisted for a long time
before the initial nucleation event, there was plenty of
time for causal
contact to be established.  The relative flatness is explained
by the automatic
``fine-tuning'' that occurs at a phase transition.
The phase surface will
grow only at the correct temperature.

Clearly much work remains to be done. A correct base theory for the
4(+1)-dimensional Euclidean space needs to be found that
would reproduce
both the standard model and General Relativity
(or generalizations of these)
on the 3(+1)-dimensional expanding phase boundary.
The model is intriguing
in its common-sense (i.e. classical)
explanations for the origin and expansion of the
universe, the source of quantum fluctuations, and the mechanism of
quantum measurement, as well as the possibility of a fully-quantum
gravitational theory. The modeling of fermions as screw dislocations
and photons as phonons is also intriguing, but not a necessary part
of the basic expansion model. For instance, another possibility could
be that the transition is more akin to that between
a normal fluid and
a superfluid, with fermions modeled as vortices -
or it could bear little
resemblance to any previously known phase transition from liquid or
solid-state physics.  Although this theory
is far from complete, which
is difficult at the outset for such a wide-ranging idea,
it is hoped that
this outline will spark further ideas that may someday form a viable
alternative to or enhancement of standard big-bang cosmology.

\end{document}